# Seasonal Variations of Extensive Air Showers with Energies from $10^{17}$ to $10^{18}$ eV

E.S.Nikiforova.
*Yu.G. Shafer Institute of Cosmophysical Research and Aeronomy,*
*31 Lenin Ave., 677980 Yakutsk, Russia*
Presenter: E.S.Nikiforova (nikiforova@ikfia.ysn.ru), rus-nikiforova-E-abs2-HE13-poster

The seasonal variations of extensive air showers (EAS) registered at the Yakutsk EAS array are considered. The dependence of variations on the atmospheric pressure is significantly expressed. The dependence on an air temperature is not found.

At the Yakutsk EAS array the ground-based stations are located in the form of grid consisting of equilateral triangles with sides of 500 m – a small "master" and 1000 m – a large "master". Showers are registered by the array when signals from three stations forming a master triangle coincide. In summer the array is usually switched off. In the present work the showers with axes located within the array perimeter with zenith angles $\vartheta < 60°$ and with charged particle densities at the "master" stations $\rho > 0.8$ particles/m$^2$ have been analysed.

Fig.1 presents the maximum and minimum temperatures during 24 hours, a mean pressure for 24 hours and the number of registered showers during 24 hours for two registration periods: from October 18, 1999 to June 14, 2000 and from September 12, 2000 to June 14, 2001. At the beginning, when switching on of the array after the summer break the gradual increase of the number of showers registered during 24 hours is seen. Then a correlation in the location of local extremums of pressure and the number of showers is seen. The number of showers increases as the pressure decreases.

Partially fluctuations of the number of showers are caused by inescapable situations when separate detectors are out of order and require repairs. During earlier registration periods so pronounced barometric dependence has not been revealed. Fig.2 presents the number of showers registered for the calendar year from 1974 to 2001. Last modernization of the array concerning the "master" stations was held in 1990 and on the plot a stable increase of statistics is seen. Registration periods, of course, have not been extended by a factor of three and the increase of statistics could be reached at the expense of the rise of reliability, stability of the array operation. At a stable work the dependence of variations on the atmospheric pressure has manifested itself.

The temperature, in contrast to the pressure, changes with a much less number of the expressed extremums in course of time. Correlation between the number of showers and a temperature in Fig.1 is not seen.

Fig.3 shows mean for several years dependences of the temperature, pressure, the number of showers on months of the year in the breakdowns of the year to 24 equal parts. The autumn increase of the number of showers is caused by a gradual switching on the array after the summer break. The decrease of pressure and the increase of the number of showers in the transition from cold months to warm ones are seen. With the rise of temperature the number of showers increases but for different ranges of the temperature the number of showers increases in different ways. Besides that, in January at constant temperature the rise of the number of registered showers takes place, while it is thought that the array evenly operates from December to May. Thus, the increase of the shower number in the transition from the cold months to warm ones is explained by a decrease of pressure but not a rise of temperature. The conclusion on average characteristics is the same as for the separate periods. As to mean characteristics from January to April the decrease of pressure of 1 mb leads to the increase of the number of showers of 1.5% or the decrease of pressure of 1 mm Hg leads to the increase of 2%.

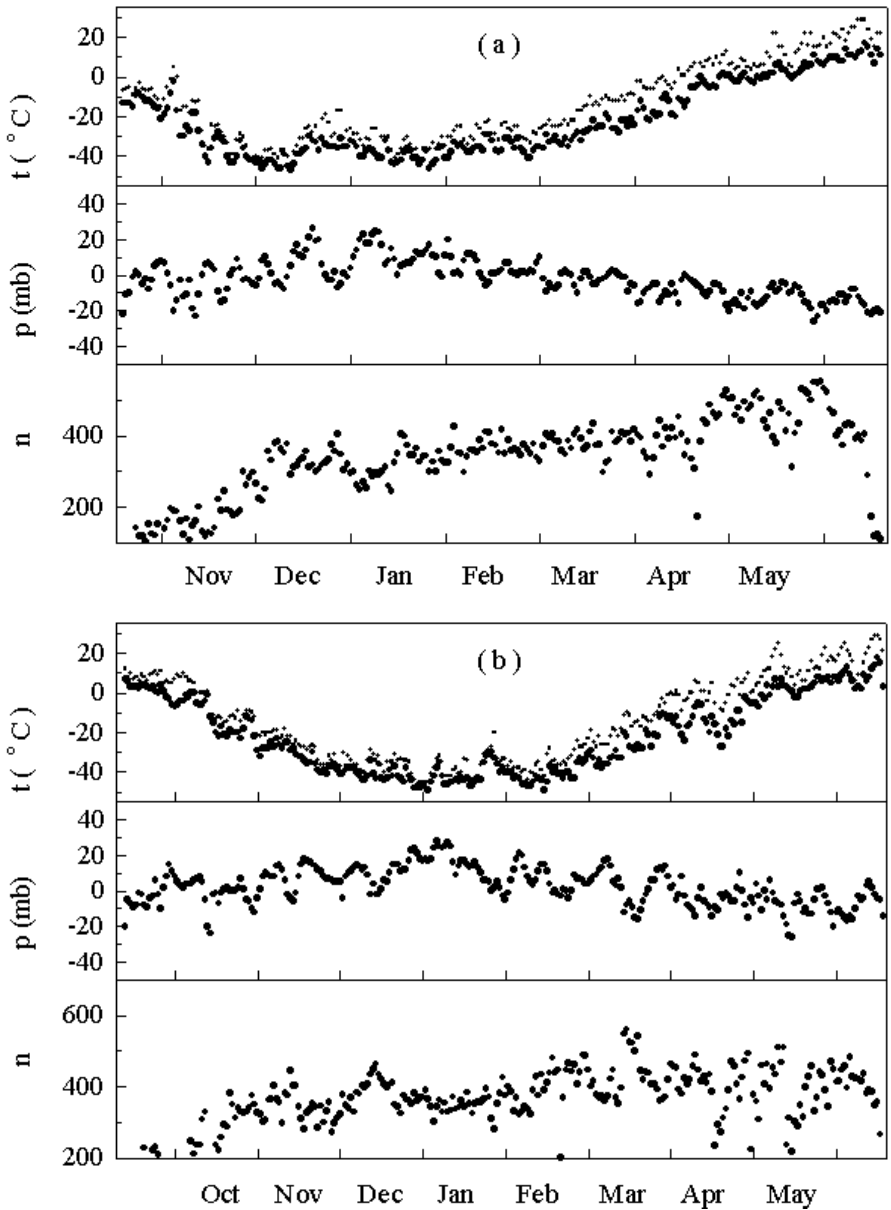

Fig.1. ● - minimum temperature, · - maximum one during 24 hours, a mean pressure for 24 hours in mb, the number of showers registered during 24 hours for two registration periods: (a) from October 18, 1999 to June 14, 2000 and (b) from September 12, 2000 to June 14, 2001.

The change of the number of showers depending on the energy is shown in Fig.4a. The black signs correspond to January and white ones - to April. It is seen that in the energy range of $10^{17}$ to $10^{19}$ eV the change of the number of showers does not depend on the energy.

At the Yakutsk EAS array in order to eliminate the influence of variations of atmospheric origin on the intensity of showers we act by the following way. The lateral distribution function (LDF) is described by the formula

$$\rho(R/R_i) \sim \left(\frac{R}{R_i}\right)^{-1} \cdot \left(1+\frac{R}{R_i}\right)^{b-1}, \qquad (1)$$

where R is a distance from the axis, $R_i$ is Moliere distance from the electro-magnetic cascade theory. It is thought that the change of temperature and pressure leads to the re-distribution of charged particles in a shower, and in this case Moliere distance changes as

$$R_i = 75 \cdot \frac{1000}{p} \cdot \frac{T}{273}, \qquad (2)$$

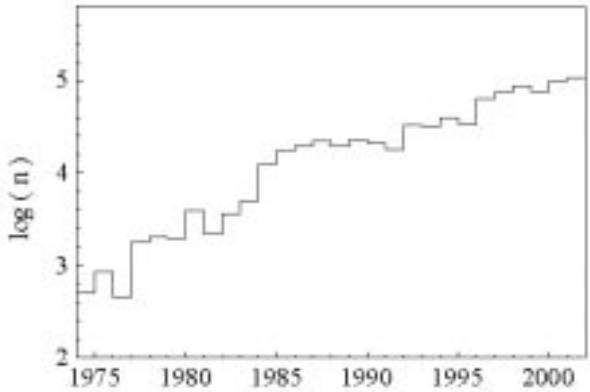

Fig.2. Distribution of showers into years.

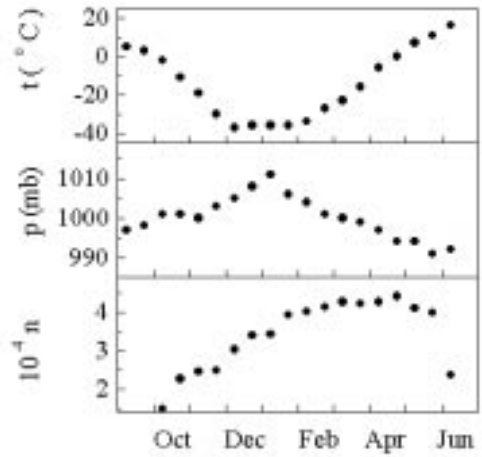

Fig.3. Dependence of the mean temperature, mean pressure, the number of showers for several years on months of the year.

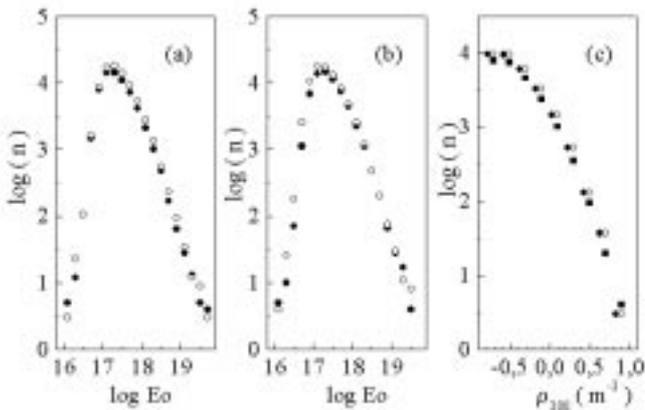

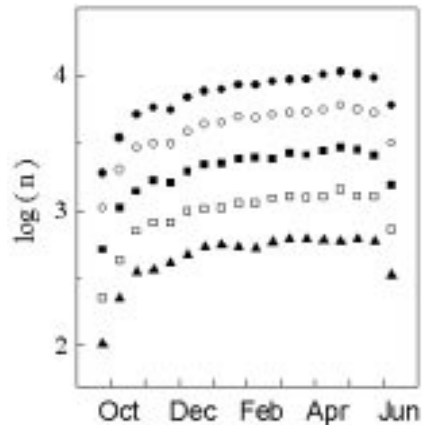

Fig.4. Dependence of the number of showers registered in January - ● and April - ○ on the energy: (a) - without recalculation of Moliere distance, (b) - with a recalculation of Moliere distance ($E_0$ is measured in electronvolts), (c) - dependence of the number of showers with $\cos\vartheta > 0.9$ registered by a small "master" on a density of charged particles at a distance of 300 m from an axis $\rho_{300}$: ■ - in January, ○ - in April, ● - in April shifted the left by 0.07 particles/m².

Fig.5. Dependence of the number of showers on months of the year:
● - $\langle\cos\vartheta\rangle = 0.95$, ○ - $\langle\cos\vartheta\rangle = 0.85$,
■ - $\langle\cos\vartheta\rangle = 0.75$, □ - $\langle\cos\vartheta\rangle = 0.65$,
▲ - $\langle\cos\vartheta\rangle = 055$.

where the temperature is in absolute degree and pressure is in mb [1]. It is also thought that the total number of particles in a shower does not change. On this basis the density of particles at a distance of 300 and 600 m for a small "masters" and a large one is recalculated by Moliere distance $R_o = 68$ m by the formula

$$\rho_*(R_o) = \rho_*(R_i) \cdot \left(\frac{R_o}{R_i}\right)^{b-2} \cdot \left(\frac{r+R_i}{r+R_o}\right)^{b-1}. \quad (3)$$

Here r=300 or 600 m for parameters $\rho_{300}$ and $\rho_{600}$, respectively, b is the parameter of LDF.

The change of the number of showers recalculated by formulae (2) and (3) depending on the energy is shown for January and April in Fig.4b. As seen from the plot at such a recalculation the number of showers at energies $E_0 > 10^{17}$ eV does not vary depending on a season. Differences at energies less $10^{17}$ eV are considered as the consequence of influence of the registration threshold. In the formula (1) Moliere distance $R_i$ depends on the temperature more strongly than on the pressure. In the transition from January to April $R_i$ changes by 14% depending on the temperature and by 2% - on the pressure. As is seen from Fig.3 the seasonal changes of temperature and pressure in some degree correlate, or, more exactly saying, anticorrelate so taking into account mainly the changes of temperature in the formulae (1) and (2), we indirectly take into account the changes of pressure.

The dependence of the number of showers registered by a small "master" for different months with a step in zenith angle $\Delta\cos\vartheta = 0.1$ is shown in Fig.5. For showers with $<\cos\vartheta>=0.55$ a seasonal dependence from January to April is less expressed, than for showers with $<\cos\vartheta>=0.95$.

The change of the number of showers depending on a density at a distance from the axis 300 m $\rho_{300}$ for a small "master" at $\cos\vartheta > 0.9$ is shown in Fig.4c. Black signs correspond to January, white ones - to April. For the coincidence of curves formed by them the signs should be shifted by 0.07 particles/m$^2$ in horizontal. One can estimate the absorption length of $\lambda$-parameter under changes of atmospheric pressure by two values of pressure using the formula

$$\rho_{300}(Apr) = \rho_{300}(Jan) \cdot \exp\left(\frac{1.02 \cdot (p_{Jan} - p_{Apr})}{\lambda}\right). \quad (4)$$

The absorption length obtained by such a way is $\lambda = 110$ g/cm$^2$ whereas the absorption length of the parameter $\rho_{300}$ obtained at fixed values of integral intensity for showers at different zenith angles is $\lambda_{300} = (310 \pm 20)$ g/cm$^2$ [2] or in the adjusted form $\lambda_{300} = (434 \pm 15) - (62 \pm 9) \cdot \log(\rho_{300}(0°))$ g/cm$^2$ [3]. The obtained values $\lambda$ for vertical showers and $\lambda_{300}$ for inclined showers differ approximately by a factor of three. So the development of vertical and inclined showers is significantly different. Decay processes begin to predominate with a decreasing density for inclined showers.

This work was partly financed by Ministry of Education and Science of Russia.